\documentclass[traditabstract]{aa}
\usepackage{graphicx}
\usepackage{amsmath}
\usepackage{txfonts}
\usepackage{natbib,twoopt}
\def\apjl{Astrophys.\ J. }
\def\apjs{Astrophys.\ J.\ (Supp.) }
\def\aap{Astron.\ Astrophys. }

\def\koz {{}}

\begin{document}

   \title{Dynamical analysis and constraints for the HD\,196885 system}

   \author{C. A. Giuppone \inst{1}
          \and
          M. H. M. Morais\inst{1}
          \and
          G. Bou\'e\inst{2,3}
          \and
          A. C. M. Correia\inst{1,2}
          }

   \institute{Department of Physics, I3N, University of Aveiro,
	      Campus Universit\'ario de Santiago, 3810-193 Aveiro, Portugal \\
              \email{cristian@ua.pt}
         \and
       Astronomie et Syst\`emes Dynamiques, IMCCE-CNRS UMR8028, 
77 Av. Denfert-Rochereau, 75014 Paris, France
       \and
       Centro de Astrof\'\i sica, Universidade do Porto, Rua das Estrelas,
	      4150-762 Porto, Portugal             }

   \date{Received September 15, 1996; accepted March 16, 1997}

% \abstract{}{}{}{}{} 
% 5 {} token are mandatory
 
 \abstract{
The HD\,196885 system is composed of {a binary star} and a planet orbiting the primary.
The orbit of the binary is fully constrained by astrometry, but for the planet the inclination with respect to the plane of the sky and the longitude of the node {are unknown}.
Here we perform a full analysis of the HD\,196885 system by exploring the two free parameters of the planet and choosing different sets of angular variables.
We find that the most likely configurations for the planet is either nearly coplanar orbits (prograde and retrograde), or highly inclined orbits near the Lidov-Kozai equilibrium points, { $ i  = 44^\circ $ or $ i  = 137^\circ $}.
Among coplanar orbits, the retrograde ones appear to be less chaotic, while 
for the orbits near the Lidov-Kozai equilibria, those around $\omega= 270^\circ$ are more reliable, {where $\omega_\koz$ is the argument of pericenter of the planet's orbit with respect to the binary's orbit}.
{ From the observer's point of view (plane of the sky) stable areas are restricted to $(I_1, \Omega_1) \sim (65^\circ, 80^\circ)$, $(65^\circ,260^\circ)$, $(115^\circ,80^\circ)$, and $(115^\circ,260^\circ)$, {where $I_1$ is the inclination of the planet and $\Omega_1$ is the longitude of ascending node}.
 }}

   \keywords{celestial mechanics --
             planets and satellites: dynamical evolution and stability -- (Stars:) binaries (including multiple): close -- (Stars:) individual: HD\,196885}
   \maketitle

\section{Introduction}

About 20\% of all known exoplanets have been found to inhabit multiple stellar systems \citep{Desidera_Barbieri_2007, Mugrauer09}. 
The theories of planet formation {around a star that is part of a} binary system have considerable challenges.
A brief review of theories can be seen in \citet{Quintana2002, Thebault_2011}. However, it is very important to know the real masses and the spatial configuration of such systems in order to better understand the processes involved in their formation. 

Among planets in binaries,
a few of them are found in compact systems with {semimajor axis of the binary} less than 100\,AU: HD\,196885 \citep{Correia_etal_2008,Chauvin_etal_2011}, Gl\,86 \citep{Queloz_etal_2000, Lagrange_etal_2006}, $\gamma$-Cep \citep{Hatzes_etal_2003, Neuhaeuser_etal_2007}, and HD\,41004 \citep{Zucker_etal_2004}. Even for these tight systems, the poor precision of first epoch observations together with incomplete time span of the observations lead to several best fit solutions with almost the same residuals (e.g \citet{Torres2007, Correia_etal_2008}). It is then advisable to keep in mind this wide range of possible configurations, and to combine radial velocity fits with possible formation scenarios \citep[e.g. $\gamma$-Cephei,][]{Giuppone_etal_2011}.

Recent  observations of the HD\,196885 system allowed to constrain the orientation of the binary orbit combining astrometric and radial velocity observations \citep{Chauvin_etal_2011}. Thus, the real mass of the stellar companion was established at 0.45 $M_{\sun}$. 
A few numerical integrations were carried by \citet{Chauvin_etal_2011}, however it is not clear which are the compatible regions of solutions for the planet around the central star and why some individual coplanar solutions are unstable. 
We intend to clarify this picture by performing massive numerical integrations over the entire space of the free parameters  
and compare the results with analytical models. This will allow us to completely clarify all the possible dynamical regimes in the HD\,196885 system, and to put constraints on the forthcoming observations.

\section{Best fit solution and reference angles}

HD\,196885\,B,
the stellar companion of HD\,196885\,A, was detected combining imaging and spectroscopic observations, as being a M1$\pm$IV dwarf located at 0.7~$\!''$, which corresponds to 23~AU in projected physical separation \citep{Chauvin_etal_2007}. 

The planetary companion, HD\,196885\,Ab, was detected using radial velocity data from ELODIE, CORALIE and CORAVEL observations spread over 14~years \citep{Correia_etal_2008}. 
The orbital solution for the planet gave a minimum mass of $m_{Ab}\,\rm{sin}i=2.96$~$M_{\rm{Jup}}$, a period of $P=3.69\pm0.03$~yr,
and an eccentricity of $e=0.462\pm0.026$. The authors additionally constrained the binary companion HD196885\,B with a
period of $40< P < 120$~yr, a semi-major axis $14 < a < 30$~AU, and a minimum mass of $0.3 < m_B \sin i < 0.6~M_{\odot}$. Based on Lick observations, \citet{Fischer_etal_2009} confirmed this solution for the system. In both studies, a relatively large range of masses and periods were found compatible with residuals for the stellar companion. 

More recently, Chauvin et al (2011) combined five astrometric measurements obtained with VLT/NACO spread over 4 years and all sources of radial velocity data (four sets, summarizing 187 observations) to determine the full orbit of the binary in the space and a consistent projection of planetary parameters. 
We show the best fit obtained by Chauvin et al. (2011) in Table~\ref{tab-chauvin}.  

\begin{table}[t]
\caption{Orbital parameters for the HD\,196885 system obtained by Chauvin et al. (2011) for JD = 2455198.}
\label{tab-chauvin} 
\footnotesize
\centering
\begin{tabular}{llcc}     % 7 columns
\hline\hline\noalign{\smallskip}
 Param.      & [unit]           & orbit 1 (planet)       & orbit 2 (binary)     \\
\noalign{\smallskip}\hline\noalign{\smallskip}
$a_i$          & [AU]             & $2.6\pm0.1$        & $21.00\pm0.86$ \\
$e_i$          &                  & $0.48\pm0.02$      & $0.42\pm0.03$ \\
$\omega_i$     & [deg]            & $93.2\pm3.0$       & $241.9\pm3.1$   \\
$M_i$          & [deg]            & $349.1\pm 1.8 0$        & $121 \pm 45$  \\ 
$\Omega_i$     & [deg]            &     ?             & $79.8\pm0.1$   \\
$I_i$          & [deg]            &      ?              & $116.8\pm0.7$  \\
$m_i$ & [$M_{\rm{Jup}}$] & $2.98/\sin{I_1} $ & $472$\\     
% $t_P$        &                  & $2002.85\pm0.02$ & $1985.59\pm0.39$\\ 
\noalign{\smallskip}\hline
\end{tabular}
\end{table}

The characterization of the orbits in the HD\,196885 system by \citet{Chauvin_etal_2011} left only two parameters undetermined, the inclination of the planet with respect to the plane of the sky, $I_1$, and the longitude of the node, $\Omega_1$ (Tab.\ref{tab-chauvin}).
It is then possible to cover the entire phase-space for this system, by exploring these two missing angles.

\begin{figure}
\centerline{\includegraphics*[width=\columnwidth]{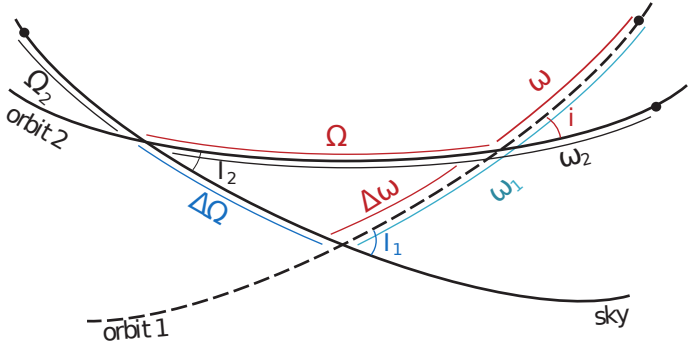}}
\caption{Fundamental planes for the definition of the reference angles.
The angles in black are determined directly from the observations (Tab.\,\ref{tab-chauvin}), while
the two angles in blue $(I_1, \Delta \Omega)$  are the missing angles in the observations that prevent the full characterization of the orbits in the system.
The angles in red  $(i, \Omega_\koz, \omega_\koz)$ are also undetermined, but related with the other angles. They are linked to the physical system, thus independent of the reference frame, and more suitable to study the dynamics.}
\label{fig-1}
\end{figure}

Before studying the phase-space of the system, it is convenient to understand all the physical configurations. A scheme of the fundamental planes for the definition of the reference angles is shown in Figure~\ref{fig-1}. For simplicity, we reserve for the inner orbit (planet) the index 1, and for the outer orbit (stellar companion) the index 2. The central star has mass $m_0$=1.31 $M_\odot$ \citep{Chauvin_etal_2011}, the planet has mass $m_1$, and the stellar companion has mass $m_2$, with semi-major axis $a_i$, eccentricity $e_i$, mean anomaly $M_i$, argument of pericenter $\omega_i$, longitude of ascending node $\Omega_i$, and inclination $I_i$. We also mark some additional angles that are useful for our study: the mutual inclination $i$, the nodal longitude of the planet's orbit  with respect to the binary's orbit 
$\Omega_\koz$, the difference of node longitudes $\Delta\Omega=\Omega_1-\Omega_2$ in the plane of the sky, the argument of pericenter of the planet's orbit with respect to the binary's orbit $\omega_\koz$, and the angle $\Delta\omega=\omega_1-\omega_\koz$. 
  
The dynamics of the system can be studied using different sets of the free parameters, the most important being ($I_1,\Delta \Omega$), ($i,\Omega_\koz$), and ($i,\omega_\koz$).
The observed quantities obtained from fits to the data are deduced with respect to plane of the sky ($I_1,\Omega_1$), but the dynamics of inclined systems is more adequately described using as reference the stellar companion (the orbit 2), since it is fully determined (Tab.\,\ref{tab-chauvin}) and it is not much disturbed by the planet.

All the reference angles are related between each other, so whatever is the choice that we adopt, one can easily determine the remaining angles by spherical trigonometry transformations \citep[e.g.][]{Smart_1965}.
If one studies the system for mutually inclined configurations ($i, \Omega_\koz$), then we get for the remaining variables:
\begin{eqnarray}
\cos{I_1} &=& \cos{I_2} \cos{i}-\sin{I_2} \sin{i} \cos{\Omega_\koz}  \ , \label{eq1}  \\
\cos{\Delta\omega} &=& \frac{(\cos{I_2} - \cos{I_1}\cos{i})}{(\sin{I_1}\sin{i})} \ , \label{eq2} \\
\cos{\Delta\Omega} &=&\cos{\Omega_\koz} \cos{\Delta\omega}+\sin{\Omega_\koz} \sin{\Delta\omega} \cos i \ , \label{eq3}    
\end{eqnarray}
noting that when $0^\circ < \Omega_\koz < 180^\circ$, we have $ 0^\circ < \Delta\omega < 180^\circ$ and $ 0^\circ <\Delta\Omega <180^\circ$, and for $180^\circ < \Omega_\koz < 360^\circ$, we have $180^\circ < \Delta\omega < 360^\circ$ and $180^\circ <\Delta\Omega <360^\circ$.

Alternatively, if one prefers to explore the space of possible solutions starting from the observer's point of view ($I_1,\Omega_1$), the spherical triangle defined by the sides $\Delta\Omega,\Delta\omega,\Omega_\koz$ has three known parameters $\Delta\Omega,I_1,I_2$ and the following relations can be used to determine the physical system:

\begin{eqnarray}
\cos{i} &=&\cos{I_2} \cos{I_1}+\sin{I_2} \sin{I_1} \cos{\Delta\Omega} \ , \label{eq4} \\
\cos{\Delta\omega} &=& \frac{(\cos{I_2} - \cos{I_1}\cos{i})}{(\sin{I_1}\sin{i})} \ , \label{eq5} \\
\cos{\Omega_\koz} &=& \cos{\Delta\Omega} \cos{\Delta\omega}-\sin{\Delta\Omega} \sin{\Delta\omega} \cos{I_1} \ . \label{eq6}
\end{eqnarray}

\section{Analytical model}

\label{helena}

Before studying the full massive problem, it is  useful to look at the restricted inner problem (the orbit of the outer companion is fixed).
This approximation is not very far from being true, since the mass of the planet is much smaller than the mass of the stellar bodies ($m_1 \ll m_0, m_2$), and can therefore be seen as a ``test particle'' ($m_1 = 0$).
The restricted problem is easier to be studied and allows us to determine the dynamical regimes that can be expected in the HD\,196885 system.

\subsection{Restricted quadrupolar problem} 

We consider a binary star system composed of a primary ($m_0$) and a secondary ($m_2$), and a massless planet ($m_1=0$) orbiting the primary star. The binary system's fixed orbit with period $T_2$, semi-major axis $a_2$ and eccentricity $e_2$, is the natural choice of reference frame. The planet's osculating orbit with period $T_1$, semi-major axis $a_1$ and eccentricity $e_1$, has orientation with respect to the binary system's orbit defined by the angles $i$ (relative inclination), $\omega_\koz$ (argument of the pericentre) and $\Omega_\koz$ (longitude of the ascending node).
Following \citet{Kozai_1962},  \citet{Kinoshita_Nakai_1999, Kinoshita_Nakai_2007}, we write the Hamiltonian of the planet, expand up to quadrupole order in the semi-major axis ratio $a_1/a_2$, and average with respect to the fast periods $T_1$ and $T_2$, obtaining the secular quadrupole Hamiltonian
\begin{equation}
F=C \left[(2+3\,e_1^2)(3\,\cos^2{i}-1)+15\,e_1^2\,\sin^2{i}\,\cos(2 \omega_\koz)\right]  
\label{hamiltonian0}
\end{equation}
with\footnote{The expression for $C$ (Eq.~7) in \citet{Kinoshita_Nakai_2007} should have $a_1^2$ and not $a_2^2$ in the nominator.}
\begin{equation}
C= \frac{{\cal G}}{16}\frac{m_2}{(1-e_2^2)^{3/2}}\frac{a_{1}^2}{a_{2}^3} \ . \label{hamc}
\end{equation}

\begin{figure*}
\begin{center}
\includegraphics[height=12 cm]{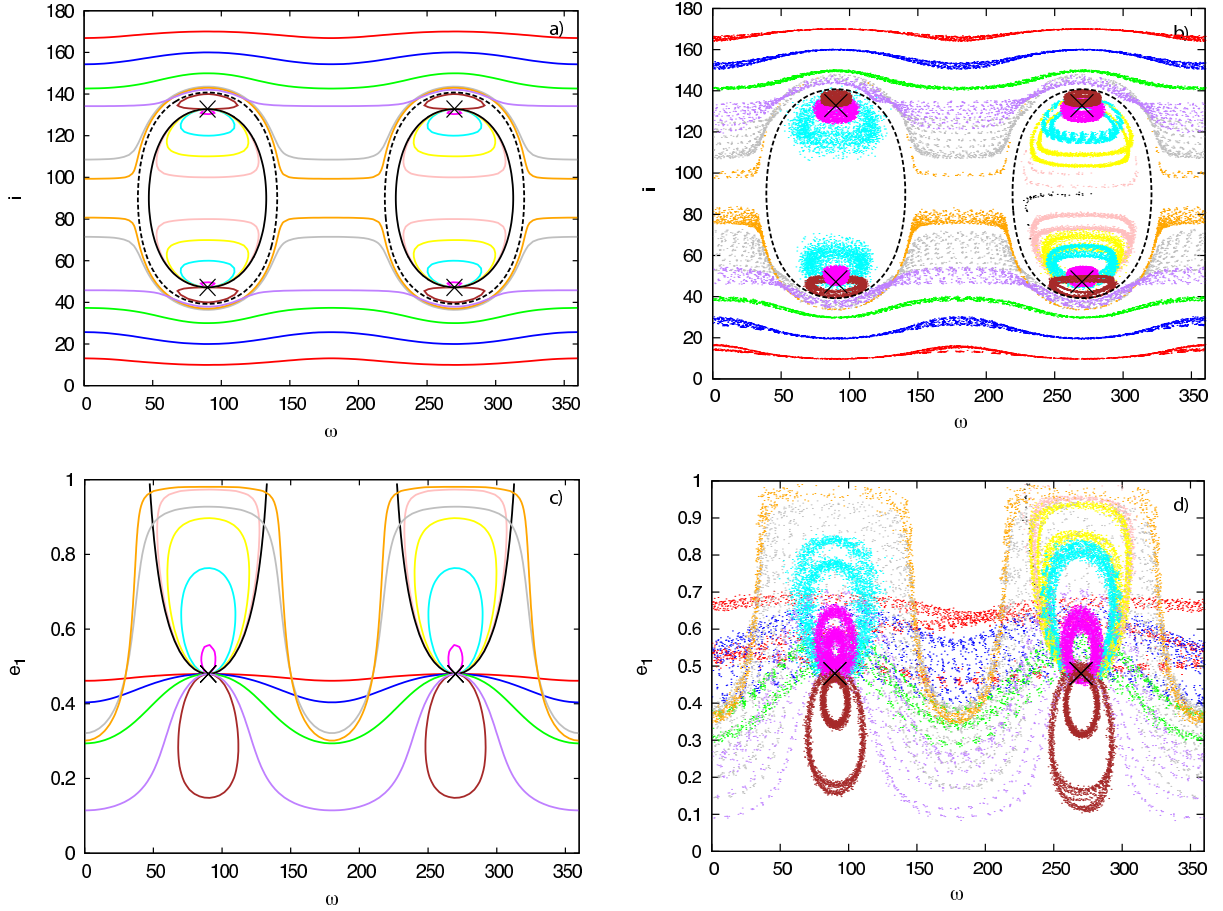} 
\caption{Possible secular trajectories for the HD196885 system seen in the
$(\omega_\koz,i)$ plane (top), and in the $(\omega_\koz,e_1)$ plane (bottom). 
We show the trajectories in the restricted quadrupolar approximation (left) and using n-body numerical simulations of the real system (right).
The dashed black curves in (a) separate zones of libration and circulation of the
angle $\omega_\koz$. The black crosses in (a) and ({c}) indicate the location of
Lidov-Kozai equilibrium points for this system.  
N-body numerical integrations in the ($\omega_\koz,i$) plane ({b}), and in the
($\omega_\koz,e_1$) plane (d), preserve the same colors for the initial conditions in the restricted problem.
\label{figkozai}}
 \end{center}
\end{figure*}

The secular motion can be described  by writing Hamilton's equations using Delaunay canonical variables  
\begin{equation}
\left( \omega_\koz, G=\sqrt{{\cal G}\,m_0\,a_1\,(1-e_1^2)} \right) \ ,  
\end{equation}
\begin{equation}
\left( \Omega_\koz, H=\sqrt{{\cal G}\,m_0\,a_1\,(1-e_1^2)}\,\cos{i} \right) \ ,
\end{equation}
where ${\cal G}$ is the gravitational constant.
Since the Hamiltonian (Eq.~\ref{hamiltonian0}) does not depend on $\Omega_\koz$, then the conjugate momentum $H$ is constant. Moreover, as $F$ (Eq.~\ref{hamiltonian0}) has only one degree of freedom, $\omega_\koz$, we obtain secular trajectories by plotting level curves $F=const$, with 
\begin{equation}
h=\frac{H^2}{({\cal G}\,m_{0}\,a_1)}=(1-e_1^2)\,\cos^2{i}=const \ .
\label{eqkozai} 
\end{equation}

In our particular case (the HD196885 system), we know the current value of the planet's eccentricity ($e_{10}=0.48$) but we do not know the initial $i_{0}$ or $\omega_{\koz{0}}$. We can obtain possible secular trajectories for the planet by choosing values of $i_{0}$ and $\omega_{\koz{0}}$, then plotting the level curves of
\begin{equation}
\label{levelcurves_i}
F(i,\omega_\koz;h)=F(i_{0},\omega_{\koz{0}};h)
\end{equation}
or 
\begin{equation}
\label{levelcurves_e1}
F(e_1,\omega_\koz;h)=F(e_{10},\omega_{\koz{0}};h)
\end{equation}
with $h=(1-e_{10}^2)\,\cos^2{i_0}$. 

We show the secular trajectories described by expressions (\ref{levelcurves_i})  and  (\ref{levelcurves_e1}) in Figure~\ref{figkozai}(a) and {(c)}, respectively. Because of symmetry in the solution space we use the same color for trajectories with relative inclination $i$ (prograde) or $180^\circ-i$ (retrograde), and for librating trajectories around $\omega_\koz=90^\circ$ or $\omega_\koz=270^\circ$.  
The separatrixes (dashed black curves in Fig.\,\ref{figkozai}(a)) mark the boundary between libration and circulation of the angle $\omega_\koz$ and are obtained by solving the implicit equation  \citep{Kinoshita_Nakai_2007}\footnote{The solution to Eq. \ref{separatrix} is independent of $h$, as we would expect, since the separatrix is an invariant curve.}
\begin{equation}
\label{separatrix}
F(i,\omega_\koz;h)=F_s=2\,C\,(3\,h-1) 
\end{equation}
with $h=(1-e_{10}^2)\,\cos^2{i_0}$ for $(\omega_\koz,i)$.

In agreement with \citet{Lidov_1961, Lidov_1962} and \citet{Kozai_1962}, the regimes of secular motion consist of Lidov-Kozai equilibrium points (with $i$ and $e_1$ fixed)
and Lidov-Kozai cycles which due to the conservation of $h$ (Eq.~\ref{eqkozai}) exhibit coupled  oscillations in $i$ and $e_1$. 
An orbit with $e_{1}=0.48$ is at a Lidov-Kozai equilibrium point  if $(i=47.2^\circ,\omega_\koz=90^\circ)$, $(i=47.2^\circ,\omega_\koz=270^\circ)$, $(i=132.8^\circ,\omega_\koz=90^\circ)$ or $(i=132.8^\circ,\omega_\koz=90^\circ)$. 
Lidov-Kozai cycles with $\omega_\koz$ librating have $h<0.6$ and $F<F_s$ (magenta, brown, cyan, yellow and pink orbits). Lidov-Kozai cycles with  $\omega_\koz$ circulating have $h>0.6$ (red, blue and green orbits) or $h<0.6$ and $F>F_s$ (grey, orange and purple orbits).
Moreover, high amplitude Lidov-Kozai cycles (grey, pink and orange orbits)
can reach $e_1$ near unity and could thus become unstable in the full problem. 
In particular, an orbit that passes through $i=90^\circ$ must, by conservation of $h$ (Eq.~\ref{eqkozai}), reach $e_1=1$ (collision orbit).  An example of this is the black orbit in Figure~\ref{figkozai},  which started at $i=90^\circ$ and $\omega_\koz=90^\circ$ (although it is not obvious from Fig.\,\ref{figkozai}(a) due to Eq.~\ref{levelcurves_i} being undefined in this region).

Note that Figure~\ref{figkozai}(a,c) are different from the standard Lidov-Kozai diagrams  that show level curves of the Hamiltonian $F$ at constant values $h=h_0$.  In our case $e_{10}$ is fixed but  we show trajectories for different values of the initial inclination $i_0$, i.e., with different $h_0$. In particular, this explains why librating orbits in the Lidov-Kozai regime do not encircle a Lidov-Kozai equilibrium point (magenta, brown, cyan, yellow and pink orbits in Fig.\,\ref{figkozai}). 
The location of the Lidov-Kozai equilibrium points in our diagrams occur at the current planet's eccentricity ($e_1=0.48$) hence circulating and librating orbits starting at $\omega_\koz=90^\circ$  nor $\omega_\koz=270^\circ$ will have eccentricity increasing (magenta, cyan, yellow and pink orbits) or decreasing (red, blue, green and purple orbits) from this initial value (Fig.\,\ref{figkozai}).  

\subsection{Numerical simulations}

In order to compare the orbital behavior in the full massive problem with the restricted problem, we performed numerical integrations up to $t=30$~kyr starting with the same initial conditions as in Figure~\ref{figkozai}(a,{c}) ({initial conditions from Table \ref{tab-chauvin} and $\omega_\koz=90^\circ$ with $i$ from $10$ to $170$ degrees, and $\omega_\koz=270^\circ$ with $i$ from $10$ to $170$ degrees}). We  show in Figure~\ref{figkozai}({b},d) these numerical integrations as well as the position of the separatrixes and the Lidov-Kozai equilibrium points calculated with the quadrupolar approximation. We used the same colors as in Figure~\ref{figkozai}(a,{c}) to facilitate the comparison between the two models.

We see that there is good agreement between the numerical integrations of the full problem (Fig.\,\ref{figkozai}{b},d) and the theoretical trajectories of the quadrupolar restricted Hamiltonian (Fig.\,\ref{figkozai}a,{c}). 
The global behavior of the planet in the HD\,196885 system is dominated by the dynamical regimes described above, i.e., Lidov-Kozai equilibrium points, librating and circulating orbits. In the full problem, the secular trajectories exhibit a drift from the secular solutions of the quadrupole Hamiltonian. 
This behavior can be explained by including octupole and higher order terms in the Hamiltonian \citep{Ford_etal_2000ApJ,Laskar_Boue_2010,Lithwick&Naoz2011}. 
The octupole Hamiltonian is not independent of $\Omega_\koz$ and thus  $h$ (given by Eq.~\ref{eqkozai}) is not a conserved quantity. Therefore, real secular trajectories (Fig.\,\ref{figkozai}c,d) drift from the theoretical curves (Fig.\,\ref{figkozai}a,{c}) due to changes in $h$.  

We also expect deviations from the restricted quadrupole model solutions if the planet's mass becomes comparable to the star's masses although, according to \citet{Farago&Laskar2010MNRAS}, the topology of the phase space should not change. 
We see that this drift is more evident for orbits in the vicinity of the separatrixes and orbits that reach $e_1\approx 1$ (nearly collision orbits). 
Note that the black trajectory in Figure~\ref{figkozai}({b},d) is the collision orbit shown in Figure~\ref{figkozai}(a,{c}) which is unstable after only $600$~yr.
Moreover, librating orbits around $\omega_\koz=90^\circ$ seem to drift more than those around $\omega_\koz=270^\circ$ which could be due to the high planet mass solutions in this region (Sect.\,\ref{cristian}, Fig.\,\ref{fig-2}).
 
 { Also, the brown orbit in Figure \ref{figkozai}d is not a double loop structure but two different initial conditions, one for ($i<90^\circ$) and the other for ($i>90^\circ$). In Figure \ref{figkozai}c both solutions coincide while in Figure \ref{figkozai}d those with $i<90^\circ$ have a bigger loop than those with $i>90^\circ$.}

\section{Dynamical Analysis}

\label{cristian}
{Several works on dynamics in close binary systems exist. Among others \citet{Rabl&Dvorak1988} and \citet{Holman1999AJ}, investigated the long-term stability of planets in coplanar circular orbits near one of the stars. Expressions of critical semimajor axis for the planet (prograde orbits) were derived in function of the mass of binary components and eccentricity of the orbit. According to these approximations prograde orbits for this system are stable with semimajor axis less than $\sim 3.8 AU$. \citet{Wiegert1997} studied the stability of hypothetical terrestrial planets in the system $\alpha$-Centauri (whose semimajor axis and eccentricity are almost the same as in our system) for several mutual inclinations. However they fixed the longitude of the node and varied the semimajor axis (in our case the semimajor axis of the planet is very well established with the radial velocity technique). The authors also noticed that highly mutual inclined orbits are unstable.}

In this section we  analyze  the dynamics and the stability of the planetary system given in Table~\ref{tab-chauvin}.
The best choice of variables to study the global dynamics is the pair ($i,\omega_\koz$), since it allows to easily identify the fixed points and the different dynamical regimes (Sect.\,\ref{helena}).
However, this choice is not adequate as $\omega_1$ imposes constraints on $\omega_\koz$, and the physical system has some restrictions in this frame (Eq.\,\ref{eq2} or \ref{eq5}).
Thus, we are left with the pair of angles ($i,\Omega_\koz$), which cover all the possible configurations, and are still independent of the observer.

\subsection{Orbital stability}\label{orbital.stability}

We constructed grids of initial conditions with 0.5$^\circ$ of resolution and each point in the grid was then numerically integrated over 30~kyr using a Burlisch-Stoer based N-body code (precision better than $10^{-12}$) using astrocentric and osculating variables. During the integrations we computed the averaged MEGNO chaos indicator $ \langle Y \rangle$: the regular orbits yield to $\langle Y \rangle \le 2$, while larger values are indicative of chaotic motion \citep{Cincotta_2000}.
The MEGNO chaos maps uses a threshold that should be applied in order to avoid excluding stable orbits that did not converge to their theoretical value or those orbits that are weakly chaotic.
Thus following \citet{Maffione_etal_2011}, the color scale shows ``stable" orbits in blue up to $\langle Y \rangle \sim 2.5$ (a particular choice based on integration of individual orbits for very long times and due to the characteristics of this system).

\begin{figure*}
 \begin{center}
   \begin{tabular}{l l}
 \includegraphics*[width=8.5cm]{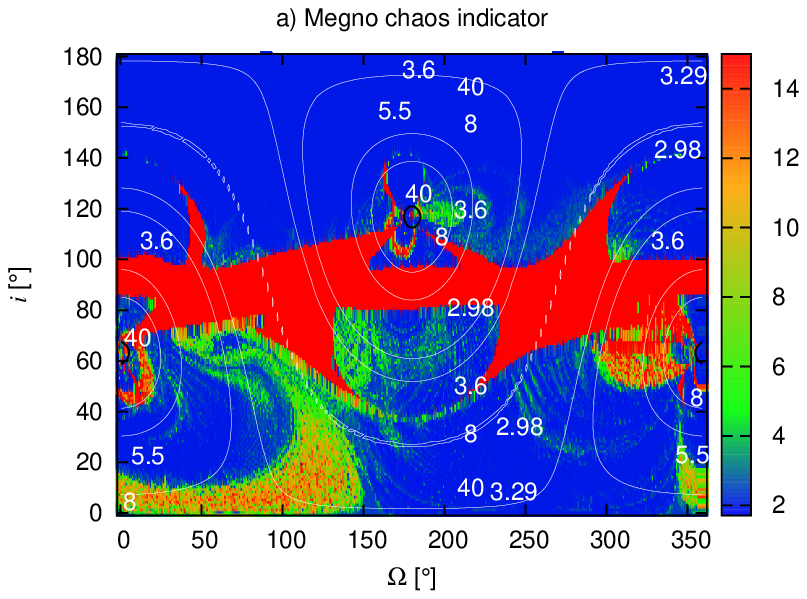} & \includegraphics*[width=8.5cm]{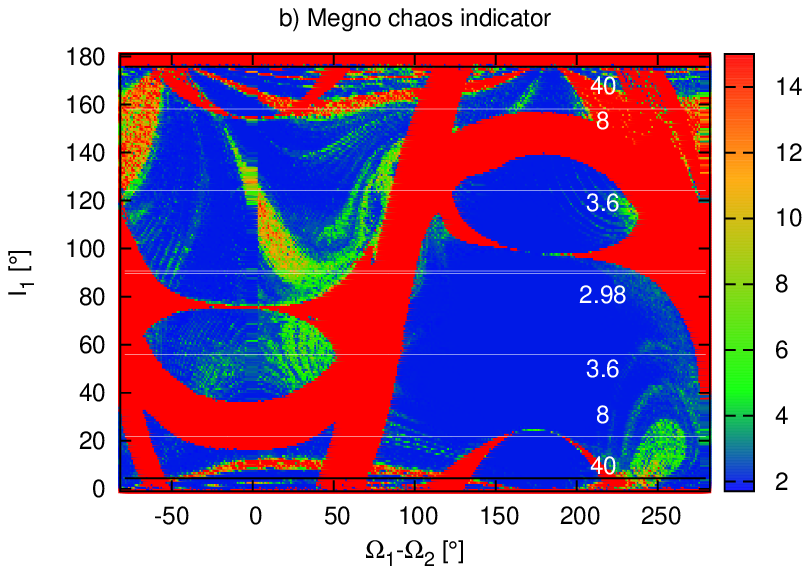}                                   \\
 \includegraphics*[width=8.5cm]{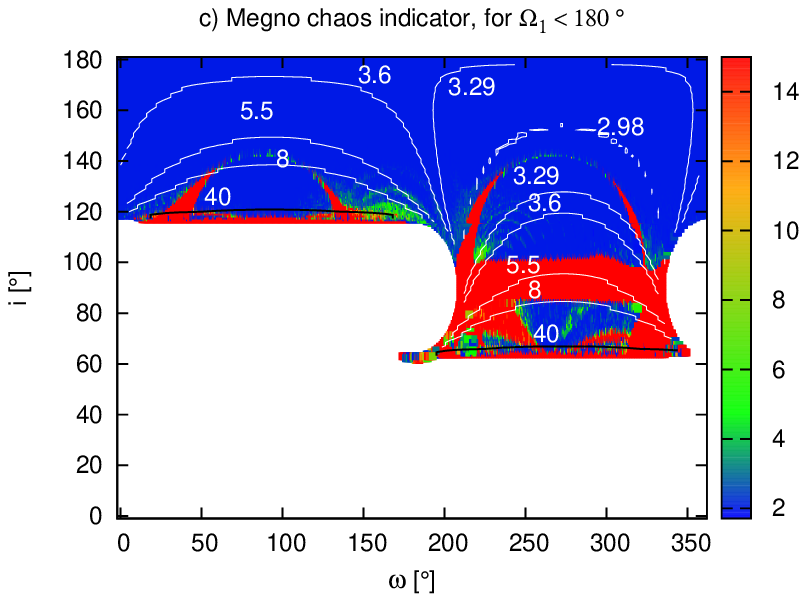} & \includegraphics*[width=8.5cm]{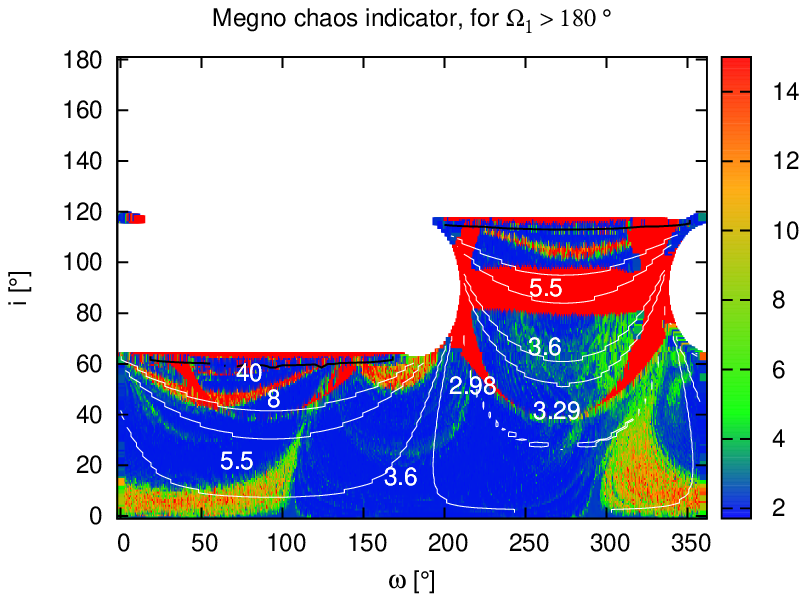} \\
 \end{tabular}
 \caption{MEGNO chaos indicators for different sets of variables, (a) ($i,\Omega_\koz$), (b) ($I_1, \Delta\Omega$), (c)  ($i,\omega_\koz$) with $0^\circ < \Omega_\koz < 180^\circ$, and (d)  ($i,\omega_\koz$) with $180^\circ < \Omega_\koz < 360^\circ$. 
Labelled lines give the mass of the planet in $M_{\rm{Jup}}$.
Orbits can be considered stable for $\langle Y \rangle \le 2$. 
 \label{fig-2}}
 \end{center}
\end{figure*}

Results are shown in Figure~\ref{fig-2}.
We also show the grids in the observational frame ($I_1, \Delta\Omega$), and in the frame  ($i,\omega_\koz$) for comparison. 
We can see that highly mutually inclined systems (the strip within $70^\circ<i<110^\circ$) are highly unstable ({this fact was already noticed in \citet{Wiegert1997} for $\alpha$-Centauri but with much less resolution)}. 

In order to better understand the chaos regions we superimposed contour level curves for mass of the inner planet (in $M_\mathrm{Jup}$). As we change the values of ($i, \Omega_\koz$) we get corresponding value of $I_1$ (Eq.\ref{eq1}) and thus the real mass of the planet ranges between 2.98 $M_\mathrm{Jup}$ (dashed lines) to 40 $M_\mathrm{Jup}$ (with larger values confined to little circles). 
Inside these circles the mass of the planet tends to infinity.
Mass level curves become straight lines in the observational frame, since $ m_1 \propto 1/ \sin I_1$, but the dynamical regimes are not perceptible, except instability corresponding to high masses for $ I_1 = 0^\circ $ and $180^\circ$.

In the frame ($i, \omega_\koz$), all the main structures become understandable: the system is unstable for high values of the mutual inclination, for high masses, but also around the separatrixes between dynamical regimes (Sect.\,\ref{helena}).
In addition, we observe that some initial conditions are not possible near high mutual inclinations ($i \sim 90^\circ$).
The reason for this ``forbidden'' zone is that some initial conditions in the frame ($i, \Omega_\koz$) are degenerate and correspond to identical solutions in the frame ($i, \omega_\koz$), although with different values for the mass of the planet.

The size of the ``forbidden'' area depends on the inclination $I_2$ of the stellar companion to the plane of the sky (Eqs.\,\ref{eq1},\,\ref{eq2}). 
For $I_2 = 90^\circ $ there is a perfect correspondence between the frames ($i, \Omega_\koz$) and ($i, \omega_\koz$), while for $I_2 = 0^\circ$ or $180^\circ$, there is a global degeneracy.
For the HD\,196885 system, we have $I_2 = 116.8^\circ$, that is, about $27^\circ$ above $90^\circ$.
As a consequence, the size of the degenerate area is about $\pm 27^\circ$ around $i=90^\circ$ in the frame ($i, \omega_\koz$).
The delimitation of this ``forbidden'' zone also coincides with the areas of large masses (Fig.\,\ref{fig-2}c,d). 
In the particular case of the HD\,196885 system, in the areas  with overlapping solutions, one always corresponds to unstable orbits (the one that is close to the borders of the zone).
Therefore, for simplicity, we can merge the two figures in a single diagram, ignoring the areas with $i < 90^\circ$ in Figure~\ref{fig-2}(c), and $i > 90^\circ$ in Figure~\ref{fig-2}(d).
For the remaining of the paper we always adopt this procedure in the frame ($i, \omega_\koz$).

{ We used the MEGNO chaos indicator because it allows us to rapidly distinguish between regular and chaotic orbits. MEGNO is very useful in these type of studies because we do not need to check every initial condition for long periods of time. We can just follow a set of orbits in representative regions of the MEGNO map to understand how chaos leads to instability. Although the MEGNO indicator only provides indications on the regularity of the orbits, we have verified that the chaotic orbits marked in red are actually unstable in the sense that the planet reaches distances to the central star smaller than the radius of the star itself (we set the limit at 0.005 AU which is the radius of the Sun, an overestimate quantity because for this system, the central star has 1.3 $R_{\rm Sun}$). The timescales of this instability depend of the chaotic region that is considered:

\begin{enumerate}
 \item Around the Kozai separatrices (red regions), chaotic orbits become unstable in short timescales (less than $\sim$ 25000 years).
 \item When the initial mutual inclination is between $80^\circ \lesssim i \lesssim 100^\circ$ the orbits are very unstable and the planet collides with the central star in less than 1000 years. As we explained in Section \ref{helena}, an orbit that passes through $i=90^\circ$ must, by conservation of $h$, reach $e_1=1$ (collision orbit).
 \item Near the edge of the forbidden region: ($i\sim60^\circ,\omega<180^\circ$) and ($i\sim118^\circ,\omega<180^\circ$), {the planets have masses $>40 M_{jup} (i_1 \sim 0^\circ)$, thus the strong interactions with both stars make the planet either collide with the central star or escape from the system. 
 \item There is a wide region of unstable orbits with high values of MEGNO inside the left-hand Kozai separatrix  ($55<i<60^\circ$ and $45^\circ \lesssim \omega \lesssim 125^\circ$) where the collision times goes from $10^5$ to $10^7$ years. }
\end{enumerate}
}

{ Within the Kozai libration islands around $\omega = 270^\circ$ the {stable} initial conditions have eccentricities that remain always lower than 0.90. Inside the Kozai libration islands around $\omega = 90^\circ$, the retrograde conditions show the same behaviour, while some prograde conditions only survive for times from $10^7$ to $10^8$ years. Outside the separatrix, in the prograde region, the chaotic orbits (in green at bottom left hand of the graph, where $\langle Y \rangle > 2.5$ and $ 4^\circ<i<10^\circ$) are stable not reaching ever eccentricities above 0.7. 

{ Long-term numerical simulations showed that the regions marked in blue/green in the MEGNO map do not present detectable dynamical instability, at least in time-scales of the order of $10^9$ years}. Due to the high mass ratio, the effect of high order mean-motion resonances (MMR, hereafter) may be non-negligible (the period ratio is 20.95). In fact, there are a multitude of high order MMR whose overlap could be responsible for the slow chaos regions. However, these high order MMR are difficult to identify, and their relative importance will depend on the exact orbital elements of the system. Even secular effects could be the explanation for the slow chaos regions in green. In particular, we refer that chaos nearby the Kozai separatrix has already been observed in the octupole problem \citep{Lithwick&Naoz2011}.

We prefer to conclude that the best representation of the system corresponds to a region of regular motion (the region marked in blue). Although this may seem arbitrary, it is important to recall that so far, in any known planetary system, there are no giant planets displaying significant chaotic motion. Secondly, regions of regular motion are expected to be more robust with respect to additional perturbations and planetary formation.
}

\subsection{Dynamical regimes and regular orbits}

\begin{figure}
 \begin{center}
   \begin{tabular}{c}
 \includegraphics*[width=8.5cm]{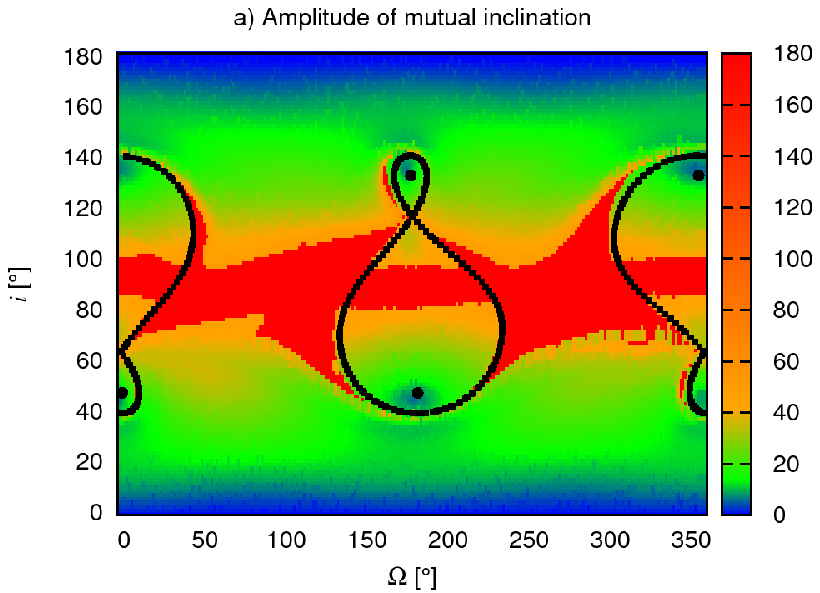}\\
 \includegraphics*[width=8.5cm]{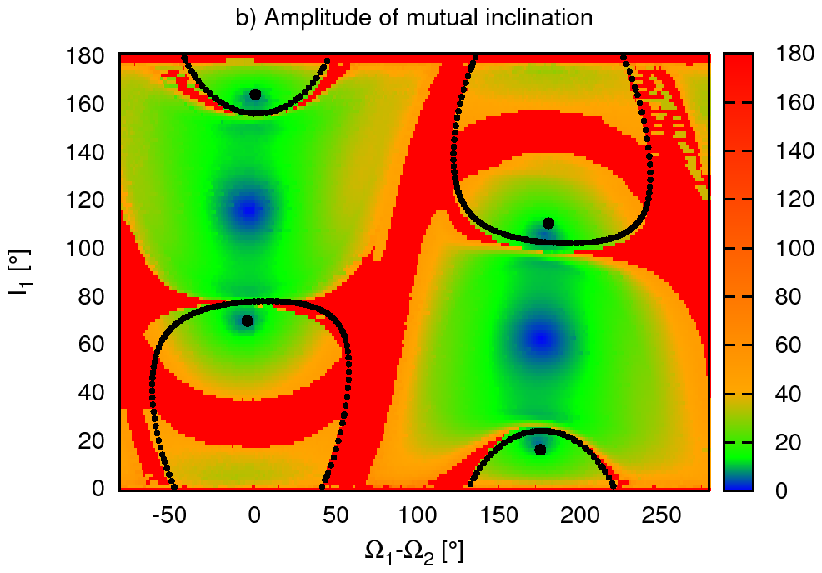}\\
 \includegraphics*[width=8.5cm]{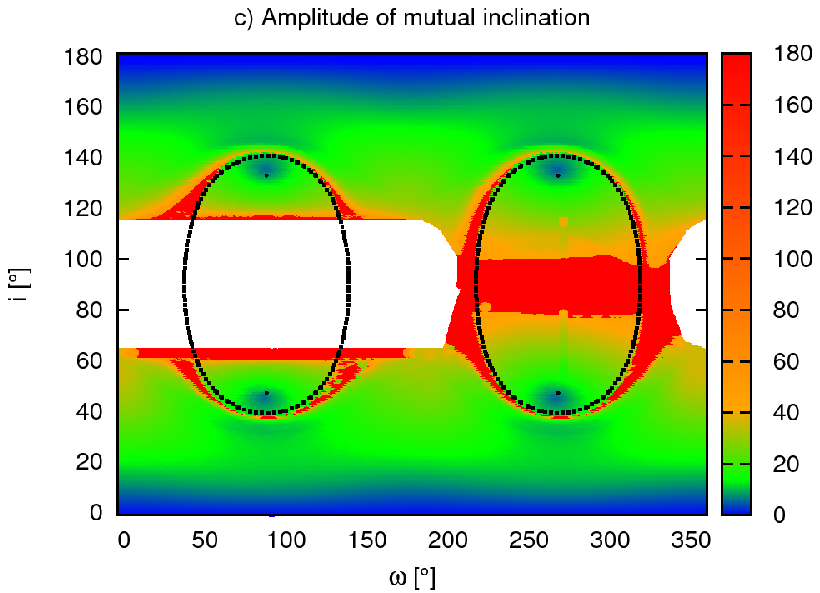} 
 \end{tabular}
 \caption{Amplitude of the mutual inclination for different sets of variables, (a) ($i,\Omega_\koz$), (b) ($I_1, \Delta\Omega$), (c)  ($i,\omega_\koz$). Centers of the Lidov-Kozai equilibria are marked as black circles and the separatrix is calculated with the quadrupolar approximation (Sect.\ref{helena}). \label{fig-12}}
 \end{center}
\end{figure}

MEGNO is a very efficient tool to identify chaotic motion, but it is not suited to distinguish between different types of regular orbits, that is, orbits that are more or less unperturbed, or orbits that experience intense secular variations. 
In order to evaluate the orbital behavior, in Figure~\ref{fig-12} we show the amplitude of the inclination variations for the three frames, together with the separatrixes between dynamical regimes obtained in Sect.\,\ref{helena}.

We observe that unstable regions (Fig.\,\ref{fig-2}) mainly coincide with the regions of large amplitude variations of the mutual inclination.
Indeed, according to expression (\ref{eqkozai}) and as discussed in Sect.\,\ref{helena}, large variations in the inclination up to $i\approx 90^\circ$ can increase the eccentricity of the inner planet to very high values, which gives rise to close encounters with the stellar companion.

The more regular orbits of the system are those which are close to the Lidov-Kozai equilibrium points, and those corresponding to coplanar orbits.
Also notorious is that retrograde orbits ($i>140^\circ$) are in general less perturbed, probably because close encounters last less time.
As we have seen before (Fig.\,\ref{fig-2}), the zones of higher mass for the planet introduce some instability, but the global picture is dominated by the dynamical regimes: trajectories that are close to the separatrix and nearly collision orbits are clearly the most unstable {(cf.\ Sect.~3)}.

\subsection{Eccentricity of the inner planet}\label{ecc}

\begin{figure}
 \begin{center}
   \begin{tabular}{c}
 \includegraphics*[width=8.5cm]{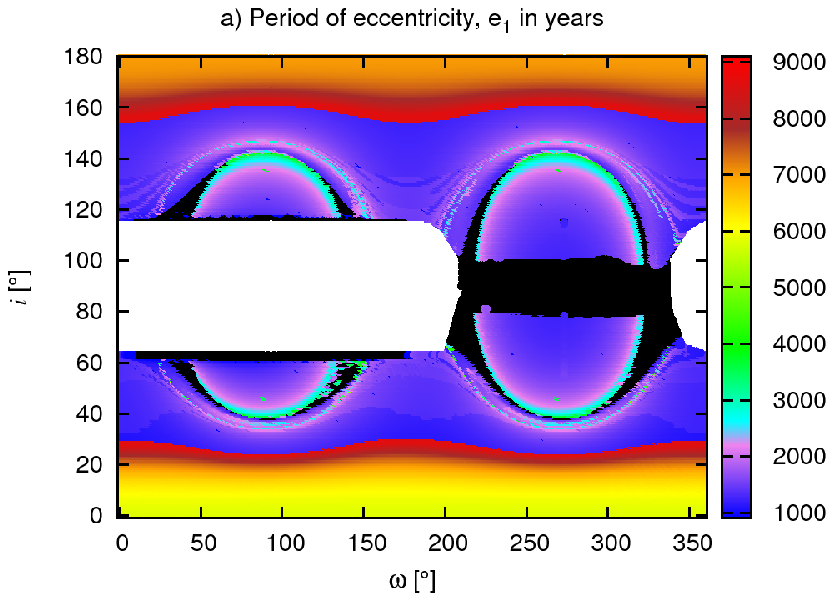}  \\
 \includegraphics*[width=8.5cm]{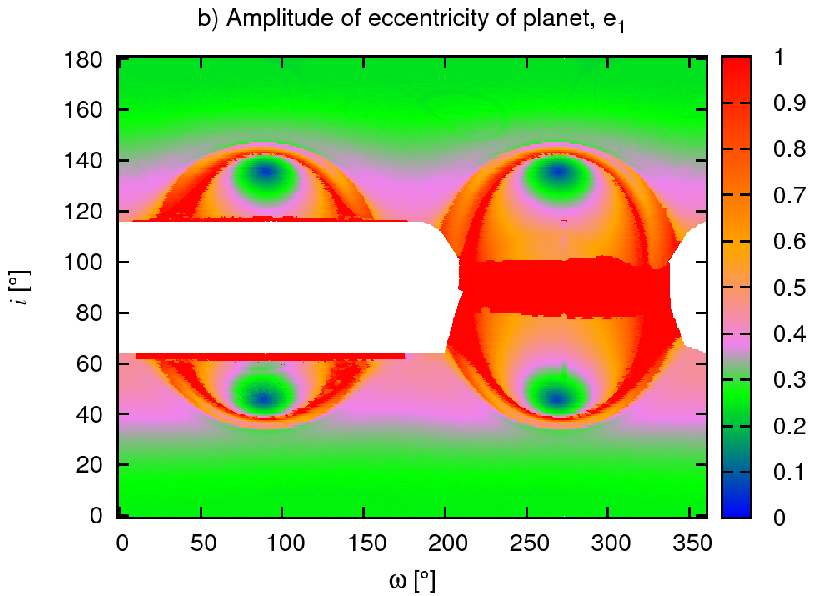} \\
 \includegraphics*[width=8.5cm]{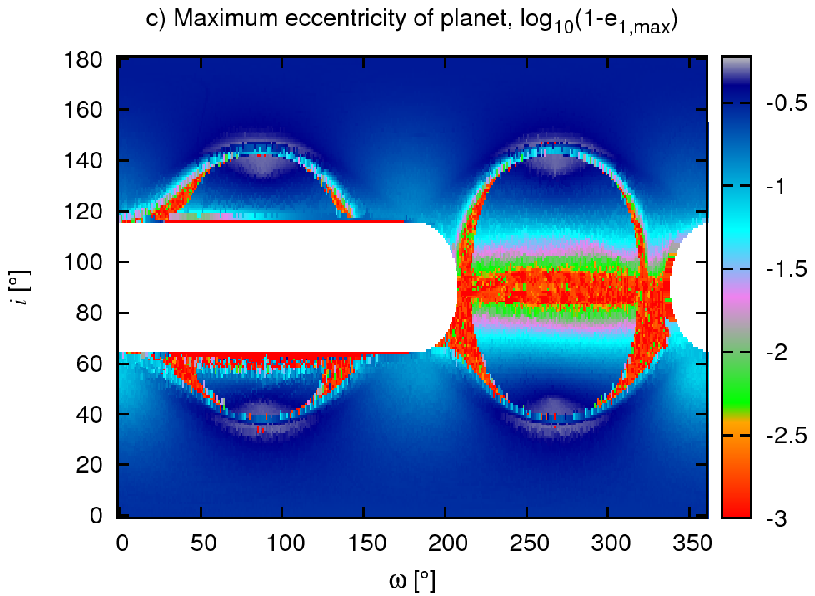} 
 \end{tabular}
 \caption{Dynamical indicators in the frame ($\omega_\koz, i$). (a) period associated to highest amplitude oscillation of $e_1$ {(in years)}, (b) maximal amplitude of oscillation of $e_1$, (c) maximum eccentricity $e_1$. \label{fig-10}}
 \end{center}
\end{figure}

The eccentricity of the inner planet is a key variable to understand the stability and the evolution of the system, since high values lead to close to collision with the central star.
Indeed, while the outer companion may destabilize the orbit of the planet, encounters with the central star may give rise to tidal effects and a subsequent secular evolution of the orbit (Sect.\,\ref{alex}).
In addition, among all orbital parameters listed on Table~\ref{tab-chauvin}, the eccentricity $e_1$ is the only one for which secular modifications can be observed using radial velocity data (the argument of the periastron $\omega_1$ also varies, but it is not directly related to the physical orbit).
In Figure~\ref{fig-10} we show the {secular period associated with the largest eccentricity oscillation of the inner planet {(in years)},  the amplitude of these eccentricity oscillations, and the maximum eccentricity, $e_1$, attained during the integration.}

\begin{figure}
 \begin{center}
  \begin{tabular}{c}
 \includegraphics*[width=8.5cm]{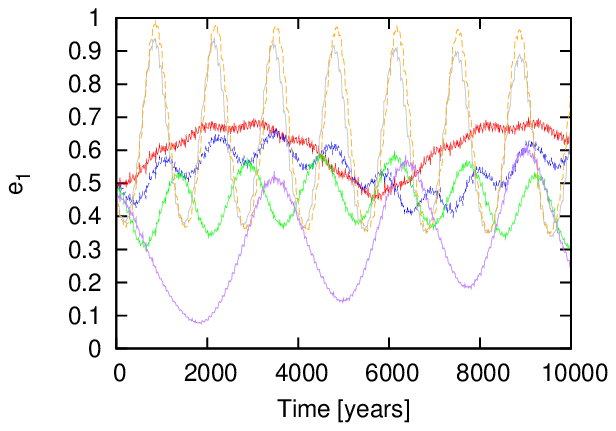} \\
\includegraphics*[width=8.5cm]{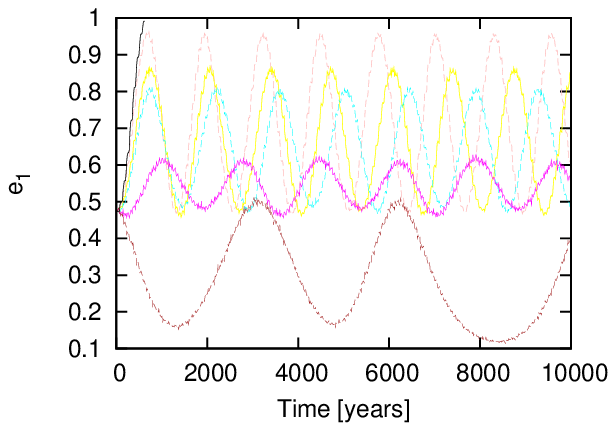}
 \end{tabular}
 \caption{Time variation of eccentricity for the orbits in Fig.
\ref{figkozai} showing the two secular periods. The color code is the
same than Fig. \ref{figkozai}. Initial
conditions outside the separatrix (top), and inside the separatrix (bottom). \label{fig4}}
 \end{center}
\end{figure}

{As discussed previously, while the behavior of high inclination orbits  (i.e.\ Lidov-Kozai cycles) is well described  with the quadrupole Hamiltonian, the secular evolution of nearly coplanar orbits must take into account the octupole Hamiltonian. In particular, the eccentrity of coplanar orbits changes secularly solely due to the octupole term \citep{Lee&Peale2003ApJ}. The eccentrity of low to medium inclination orbits is modulated by the octupole and quadrupole secular periods \citep{Krymo&Mazeh1999MNRAS}. Moreover, the quadrupole secular period is typically shorter than the octupole secular period (Fig.~\ref{fig4}). Octupole and higher order terms are also responsible for making the orbits near the separatrixes chaotic \citet{Lithwick&Naoz2011}. 

In  Fig.\,\ref{fig-10}(a)  we see the secular period associated with the largest eccentricty, $e_1$, variations.}   
From Fig.\,\ref{fig-10}(a) we see that retrograde orbits have secular periods longer than direct orbits. 
{Among stable areas, medium amplitude  Lidov-Kozai cycles have the largest eccentricity variations (Fig.~\ref{fig4} and Fig.~\ref{figkozai}c,d) and periods of around 1000~yr (Fig.~\ref{fig4} and Fig. \ref{fig-10}(a).  The eccentricity oscillations of small amplitude Lidov-Kozai cycles occur on a 1500~yr timescale while nearly coplanar orbits also exhibit small oscillations in $e_1$ but on a timescale of around 5500~yr (Fig.~\ref{fig4}).  Although we may be able to measure  eccentricity variations in the radial velocity data (for instance, by detecting a drift in the data), in practice, we will have only a short observation timespan (typically a few years). Therefore, it will be difficult to distinguish small amplitude Lidov-Kozai cycles from nearly coplanar orbits solely based on the different secular timescales (e.g.\ compare magenta and red orbits in Fig.~\ref{fig4}).  We may, however, be able to identify if the planet is on a medium amplitude Lidov-Kozai cycle due to the large eccentrity variations on a short (1000~yr) timescale.}

The amplitude of the eccentricity oscillations (Fig.\,\ref{fig-10}b) does not differ much from the amplitude of the mutual inclination oscillations (Fig.\,\ref{fig-12}c).
This behavior was expectable since these two variables are correlated (Eq.\,\ref{eqkozai}).
As for the inclination, we can easily identify the different dynamical regimes. 

{ In Figure \,\ref{fig-10}c, we plot the quantity $\log_{10}(1.-e_{1,max})$, where $e_{1,max}$ is the maximal excentricity attained by the planet during the integration. Thus a value of $\log_{10}(1.-e_{1,max})=-3$ (in red) represents an eccentricity of $e_{1,max}$=0.999. Light-blue regions are those with $e_{1,max} \sim$ 0.9, dark-blue for those with $e_{1,max} \sim 0.7$, and finally light-gray for those where $e_{1,max} \sim 0.5$. The initial conditions within violet region inside the Kozai-Lidov separatrix can survive for times from $10^8$ to $10^9$ years. Note that this confirms that the orbits marked in red in MEGNO map collide with the central star because $e\sim1$ (as described in Subsection \ref{orbital.stability})}

\subsection{Different initial conditions}

{
Since the orbital parameters in Table\,\ref{tab-chauvin} contain some undeterminations \citep{Chauvin_etal_2011}, we also explored the stability of the system within one $\sigma$ of the best fit parameters. In particular, we wanted to test the impact of variations in the outer companion, which is the most unconstrained. For that purpose, we have made grids at the upper limits $a_2=21.86$ AU and $e_2=0.45$, and at bottom ones $a_2=20.14$ AU and $e_2=0.39$ as in Figure\,\ref{fig-2}.
We do not show the results, since unstable (red regions) are located in same positions ($i \sim 90^\circ $, and around the Kozai separatrix). Only regions of moderate chaos (green regions) varied their positions and size a little bit. However, those regions are still possible solutions. Therefore, we conclude that the global picture in Figure\,\ref{fig-2} does not change much for different sets of initial conditions, that is, all the conclusions in this paper are still valid within the errorbars of the published parameters.

We also make a grid for the restricted case (i.e. the mass of the planet $m_1=0$) using the best-fit parameters from Table\,\ref{tab-chauvin}. All the chaotic green regions disappeared and we were left only with regular regions (blue) and close encounter regions (in red).}

\subsection{Tidal evolution} 
\label{alex}

For some orbital configurations, the eccentricity of the inner planet may reach very high values, and become  close enough to the central star at periastron to undergo tidal effects.
For an unperturbed orbit,
the secular evolution of the eccentricity by tidal effect can be given by \citep{Correia_2009}:
\begin{equation}
\dot e_1 = - K f (e_1) \, e_1 \ , \label{100210b} 
\end{equation}
with
\begin{equation}
f (e) = \frac{1 + \frac{45}{14}e^2 + 8e^4 + \frac{685}{224}e^6 + \frac{255}{448}e^8 + \frac{25}{1792}e^{10}}{\left(1 + 3e^2 + \frac{3}{8}e^4\right) (1-e^2)^{-3/2}}  \ , \label{090527a}
\end{equation}
and
\begin{equation}
K =  \frac{21\pi}{T_1} \frac{k_2}{Q} \frac{m_0}{m_1} \left(\frac{a_1}{a_0}\right)^3 \left(\frac{R}{a_0}\right)^5 \ ,
\label{100210c} 
\end{equation}
where $k_2 $ is the second Love number, $Q$ is the tidal dissipation factor, $R$ is the radius of the planet, and  $ a_0 = a_1 (1-e_1^2) = Const. $

The solution of the above equation is given by \citep{Correia_Laskar_2010B}
\begin{equation}
F(e_1) = F(e_{1,initial}) \exp(- K t) \ , \label{100210d} 
\end{equation}
where $ F(e) $ is an implicit function of $ e $, which converges to zero as $ t
\rightarrow + \infty $.
The characteristic time-scale for fully dampening the eccentricity of the orbit
is then $ \tau \sim 1 / K $.
{The above expression is only valid for unperturbed orbits, but in the HD\,196885 system the initial eccentricity can be seen as the maximal eccentricity attained on a Kozai cycle, $e_{1,max}$, since it is at this point that dissipation is maximized. Therefore, $\tau$ provides a minimal estimation of the dampening time.} 
For the orbits for which the eccentricity can be damped during the age of the system, we can exclude them from the possible orbital parameters of the planet, since the present eccentricity is still near 0.5 (Tab.\,\ref{tab-chauvin}).

\begin{figure}
\begin{center}
\includegraphics*[width=8.5cm]{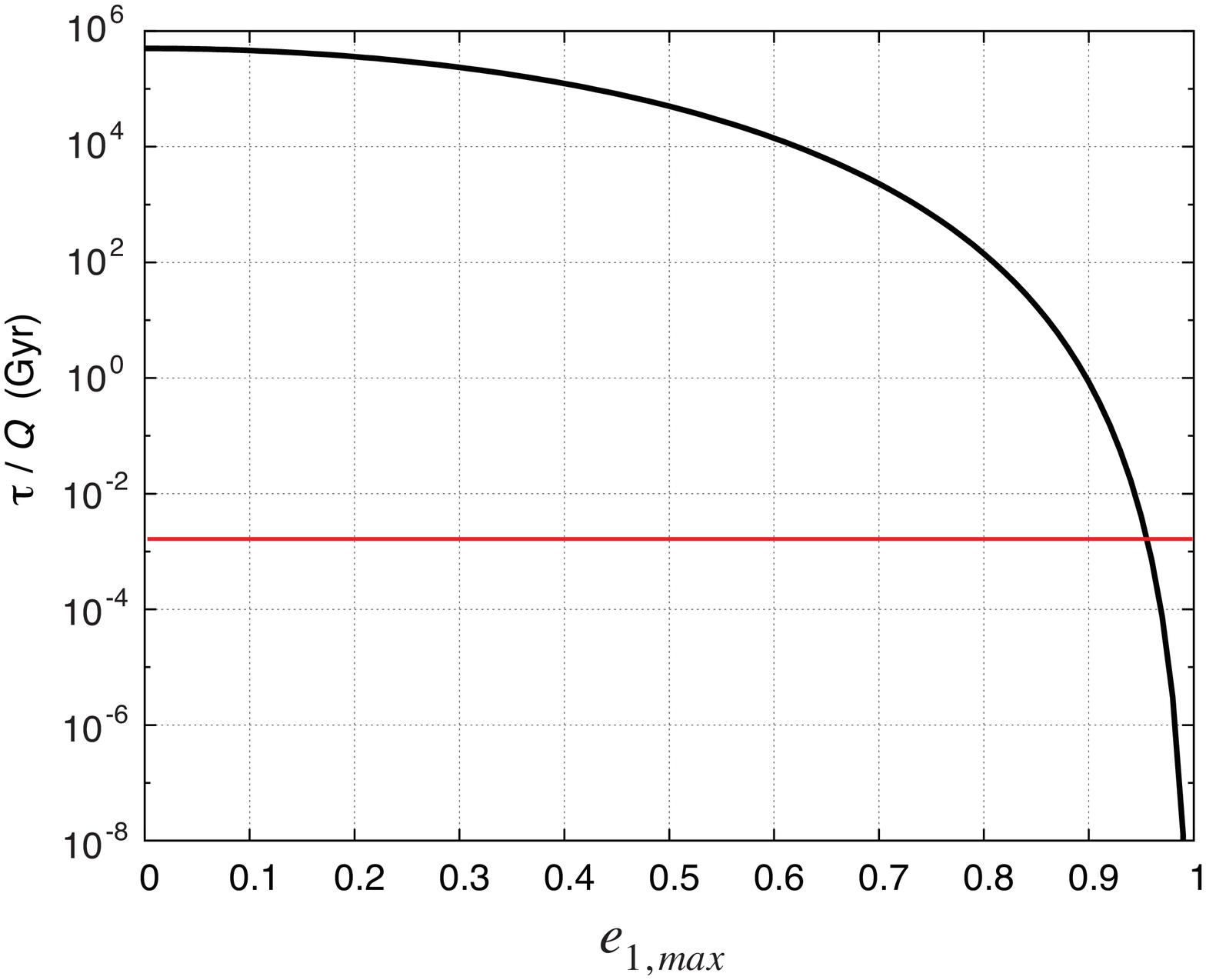} 
\caption{{Damping time for the eccentricity of the inner planet (with $\tau = 1/K $, Eq.\,\ref{100210c})}. For $Q \sim 1000$, only for $e_{1,max} > 0.96 $  the eccentricity is damped within 2~Gyr, the estimated age of the system ($\tau/Q = 2 \times 10^{-3} $~Gyr, red line). \label{figtides}}
\end{center}
\end{figure}

In Figure~\ref{figtides} we show the evolution time-scale for different initial maximal eccentricities, adopting $k_2 = 0.5$ and $ R= 1.2\,R_\mathrm{Jup}$.
We observe that for $Q \sim 1000$, a value similar to Jupiter, only for $e_{1,max} > 0.96 $  the eccentricity is damped within 2~Gyr, the estimated age of the system \citep{Correia_etal_2008}.
Even for Earth-like planets ($Q \sim 10$), only $e_{1,max} > 0.92 $ can be dissipated during the same amount of time.
Since $ e_1 > 0.96 $ corresponds to unstable orbits, we conclude that tidal effects cannot be used to exclude any of the stable initial conditions compatible with the observational data (Tab.\,\ref{tab-chauvin}).

\section{Formation hypothesis}
\label{gwen}

From the formation point of view, the orbital excitation exerted by Lidov-Kozai cycles implies destructive, high-velocity collisions among planetesimals, inhibiting formation of massive objects \citep{Lissauer_1993}. However the Lidov-Kozai equilibrium gives itself a protection mechanism to get stable configurations when mutually inclined bodies are considered. 

\citet{Thebault_2011} published a deep study of the possible scenarios of formation for the planet in the HD\,196885 binary system. 
The author considered axisymmetric static gas disc (no self gravity) and estimated the impact velocities amongst a population of circumprimary planetesimals. 
A main conclusion was that the circumprimary disc is strongly hostile to planetesimal accretion, especially the region around 2.6~AU (the planet's present location) where the binary perturbations induce planetesimal-shattering velocities of more than 1~km/s. 
The region around 2.6~AU is strongly hostile to planetesimal accretion, even for highly inclined orbits and alternative solutions were proposed to justify their existence (e.g., different initial configuration of binary or disk instability). { \citet{Quintana2002} studied the formation in $\alpha-Centauri$ system using a disk of combined large and small bodies. They concluded that it is possible to form planets when an inclined disk is considered, although $\sim 95\%$ of the initial mass is lost by the end of the simulation. Their results predict a wider region to form more retrograde orbits ($i=180^\circ$) than prograde ones and that maybe large planetary embryos could be form near the star (from 0.5 to 1.5 AU) in inclined orbits. However as the initial semimajor axis distribution has an upper limit set at $\sim$2 AU, this may bias the final results for the allowed regions. }

However, \citet{Batygin_etal_2011} address the formation for wide binary systems (a$_{\rm binary} \sim$ 1000 AU), concluding that is possible to form a single planet in an inclined orbit (Lidov-Kozai regime), if taking into account the self gravity of proto-planetary disk (that means planetesimals embedded in the gaseous disk). They also pointed out that the evolutionary process of formation of a planetary system at the Lidov-Kozai equilibrium is necessarily non-unique and proposed a scenario were a multiple system could be formed protected by Lidov-Kozai cycles and subsequent instabilities remove all the remaining planets. 
Unfortunately no mention to close binary systems with this approach could be found.

  The secular dynamics of small planetesimals play a fundamental role in establishing the possibility of accretional collisions in such extreme cases. The most important secular parameters are the forced eccentricity and the secular frequency, which depend on the initial conditions of the particles, as well as on the mass and orbital parameters of the secondary star \citep[e.g.][]{Thebault_2011, Beauge_etal_2010}. However \citet{Giuppone_etal_2011} pointed out that for these kind of compact systems, sometimes the frequencies are not quite well determined from first order approximations (second order on mass are needed), and thus probably, some works on formation based in these approximations should be reviewed. 

As for the giant planets in the Solar System, that are supposed to have migrated owing to their
interaction with a disk of planetesimals \citep[e.g.][]{Tsiganis_etal_2005}, we may assume that the same occured with the orbit of the planet in the HD\,196885 system.
Using the secular model described in \citet{Correia_etal_2011} {for the full system (two stars and one planet)}, we have performed some simple experiments on the early evolution of the system, by migrating hypothetical initial orbital parameters of the planet into the present ones.

\begin{figure}
\begin{center}
\includegraphics*[width=8.5cm]{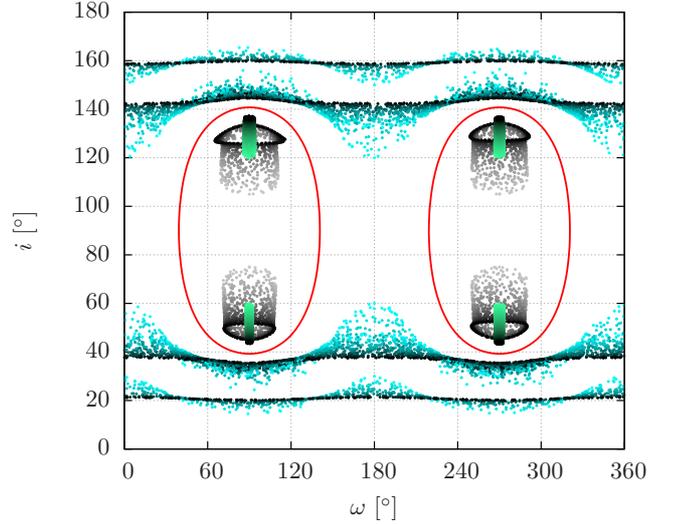} 
\caption{Early evolution of the planet, when the eccentricity is damped from higher values due to the interaction with a disk of planetesimals. For each simulation, the color of the position of the planet becomes darker with time. In the circulation zone only the amplitude of the mutual inclination is damped, but for orbits starting in the libration area, the planet evolves into the Lidov-Kozai equilibria. \label{figmig}}
\end{center}
\end{figure}

We have independently tested the evolution of the semi-major axis and the eccentricity of the inner orbit, using an exponential decay  \citep{Beauge_etal_2006, Lee_etal_2007}:
\begin{equation}
\dot a_1 = \frac{\Delta a_1}{\tau_m} \exp(-t/\tau_m) \ , \quad  \dot e_1 = - \frac{e_1}{\tau_m / 10} \ ,
\label{110912a} 
\end{equation}
with $ \tau_m = 10$\,Myr.
For the semi-major axis, there is no variation in the orbital configuration, since the ratio $a_1/a_2$ is a factor in the quadrupolar Hamiltonian (Eq.\,\ref{hamc}), so it only modifies the time-scale of the evolution.
However, for the eccentricity we observe that all trajectories which begin inside the libration zone migrate into one of the Lidov-Kozai equilibrium positions (Fig.\,\ref{figmig}).
Trajectories in the circulation zone remain more or less unchanged, only the amplitude of the mutual inclination is damped. 
Therefore, we can assume that if the planet in the HD\,196885 was able to form in libration, it will be presently observed near the Lidov-Kozai equilibria.

\section{Discussion and conclusions}

We have studied the dynamics of the  HD\,196885 system using the present determination of the orbital parameters (Tab.\,\ref{tab-chauvin}) as starting point.
We developed an insight full analysis of the 3-D space exploring the free parameters and choosing different sets of angular variables.
We found that the most likely configurations for the planet in the HD\,196885 system is either nearly coplanar orbits (prograde and retrograde), or highly inclined orbits near the Lidov-Kozai equilibrium points, $ i - 90^\circ = \pm 47^\circ $.

{Among coplanar orbits, the retrograde ones appear to be less chaotic, possibly because close encounters last less time than in the prograde case. }% \footnote{The overlap of MMR's is also less widespread in the retrograde case \citep{Moraisetal2012}.}.
For the orbits at the Lidov-Kozai equilibrium, those around $\omega= 270^\circ$ are more reliable than those around $90^\circ$, since the mass of the planet is smaller in the first situation.
Present formation scenarios for this kind of systems are unable to distinguish between the two possibilities (coplanar or Lidov-Kozai equilibrium).
However, a simple simulation on the evolution of the initial eccentricity of the inner orbit shows that if the planet was able to form in the libration zone, it will evolve into the Lidov-Kozai equilibrium point.
We also tested the effect of tides, but they are too weak in the HD\,196885 system, although they should be taken into account for tighter systems of this kind.

Although there is a wide variety of stable initial conditions for the planet in the natural frame defined by the orbit of the two stars ($i, \Omega_\koz$), from the observer's point of view (plane of the sky) there are some restrictions.
Indeed, looking at Figure~\ref{fig-12}(b) we see that stable areas {occur around:}
 \begin{center}
   \begin{tabular}{l c c}
& $\Omega_1 \sim 80^\circ$ & $\Omega_1 \sim 260^\circ$ \\
$I_1 \sim 65^\circ$  & kozai prograde & coplanar retrograde \\
$I_1 \sim 115^\circ$ & coplanar prograde& kozai retrograde
 \end{tabular}
 \end{center}
Thus, by continuing to observe the system one will be first able to determined the dynamical regime of the planet (coplanar or Lidov-Kozai equilibrium) and later its exact position.
{Moreover, it may be possible to identify medium amplitude Kozai cycles due to their large eccentricity oscillations on 1000~yr timescale. Small amplitude Kozai cycles and nearly coplanar orbits have small eccentricity oscillations on distinct timescales (1500~yr and 4000~yr timescales, respectively). However, in practice, it may  be difficult to distinguish these two configurations as we are limited to short observation timespans.}

\section*{Acknowledgments}
We would like to express our gratitude to the referee for his detailed analysis of our paper and his comments and suggestions that helped to improve the manuscript significantly. This work has been supported by the European Research Council/European Community under the FP7 through a Starting Grant. We also acknowledge financial support from FCT-Portugal (grants PTDC/CTE-AST/098528/2008 and PEst-C/CTM/LA0025/2011). 

 \bibliographystyle{aa}
 \bibliography{library}

\end{document}